\documentclass[a4paper,11pt]{article}
\usepackage{pos}
\usepackage{listings}
\usepackage{wrapfig,lipsum,booktabs}

\title{Democratizing LHC Data Analysis with ADL/CutLang}

\author*[a]{Sezen Sekmen}
\author[b]{Gokhan Unel}
\author[c]{Harrison B. Prosper}
\author[d]{Aytul Adiguzel}
\author[e]{Burak \c{S}en}

\affiliation[a]{Center for High Energy Physics, Kyungpook National University, \\
  Daegu, South Korea}

\affiliation[b]{Department of Physics and Astronomy, University of California at Irvine, \\
Irvine, CA, USA}

\affiliation[c]{Department of Physics, Florida State University,\\
Tallahassee, FL, USA}

\affiliation[d]{Physics Department, Istanbul University,\\
Istanbul, Turkey}

\affiliation[e]{Department of Physics, Middle East Technical University,\\
Ankara, Turkey}

\emailAdd{ssekmen@cern.ch}

\abstract{Data analysis at the LHC has a very steep learning curve, which erects a formidable barrier between data and anyone who wishes to analyze data, either to study an idea or to simply understand how data analysis is performed. To make analysis more accessible, we designed the so-called Analysis Description Language (ADL), a domain specific language capable of describing the contents of an LHC analysis in a standard and unambiguous way, independent of any computing frameworks. ADL has an English-like highly human-readable syntax and directly employs concepts relevant to HEP. Therefore it eliminates the need to learn complex analysis frameworks written based on general purpose languages such as C++ or Python, and shifts the focus directly to physics. Analyses written in ADL can be run on data using a runtime interpreter called CutLang, without the necessity of programming. ADL and CutLang are designed for use by anyone with an interest in, and/or knowledge of LHC physics, ranging from experimentalists and phenomenologists to non-professional enthusiasts. ADL/CutLang are originally designed for research, but are also equally intended for education and public use. This approach has already been employed to train undergraduate students with no programming experience in LHC analysis in two dedicated schools in Turkey and Vietnam, and is being adapted for use with LHC Open Data. Moreover, work is in progress towards piloting an educational module in particle physics data analysis for high school students and teachers. Here, we introduce ADL and CutLang and present the educational activities based on these practical tools.}

\FullConference{%
  *** The European Physical Society Conference on High Energy Physics (EPS-HEP2021), ***\\
  *** 26-30 July 2021 ***\\
  *** Online conference, jointly organized by Universität Hamburg and the research center DESY ***
}

\begin{document}
\maketitle

\section{Introduction}

Experiments at the CERN Large Hadron Collider (LHC) have collected an unprecedented amount of data that inspire a large diversity of physics studies. The more physics ideas we explore with these data, bigger the LHC physics output would become.  Yet, LHC data are complex, as they host large amounts of information organized in a complex structure, and data analysis is a highly non-trivial task.  It is thus essential to provide accessible analysis means to enable interested researchers or enthusiasts from all technical backgrounds to contribute to exploring these data.

At the LHC, we traditionally perform data analysis using dedicated analysis software frameworks that are written using general purpose languages such as C++ or Python.  These frameworks are detailed, usually flexible, and can handle many analysis tasks, but their inclusive and flexible nature often leads to a steep learning curve.  A leading reason is that, in traditional frameworks, the analysis physics content and technical operations are intertwined and handled together.  This introduces an ambiguity to the physics content description, and makes the analysis code hard to read, understand, modify, maintain and communicate.  

This contribution is about a new, alternative approach that allows a more direct interaction with data via decoupling the physics information in an analysis from purely technical tasks and framework infrastructure.  This approach introduces ``Analysis Description Language (ADL)", a domain specific and declarative language that describes the physics content of a collider analysis in a standard and unambiguous way~\cite{Brooijmans:2016vro}.  Here, \emph{domain specific} refers to the fact that ADL is customized to express analysis-specific concepts and reflects the conceptual reasoning of particle physicists; and \emph{declarative} means that ADL directly communicates what to do in an analysis, but not how to do it.  ADL is a language, defined independently of software frameworks.  It is based on easy-to-read, clear, self-describing syntax rules. Yet ADL is by definition a language, therefore it needs to be rendered executable in order to process the analysis on events.  As ADL is framework-independent, this can be achieved by any framework that can parse and understand the ADL syntax.  Currently, two such frameworks exist that can execute analyses written with ADL syntax at different levels: {\tt CutLang}~\cite{Sekmen:2018ehb, Unel:2019reo, Unel:2021edl} and {\tt adl2tnm}~\cite{Brooijmans:2018xbu}.  The distinction from traditional frameworks is that both CutLang and adl2tnm are automated, and hence the user does not interact directly with these frameworks or do any programming.  The user only defines the physics algorithm via ADL, while the frameworks perform all technical tasks to implement the algorithm and process it on events.  This is what makes ADL a very practical and efficient method, particularly eligible for education and outreach.

ADL is a common language intended for use by analysts with a wide range of technical knowledge and skill for a wide spectrum of projects.  ADL and tools can be used for experimentalists for real data analysis, phenomenologists for model sensitivity or reinterpretation studies as well as by students or interested public for learning HEP data analysis and creating analysis projects with public data or Monte Carlo (MC) events.  This allows the focus to be directly on physics, rather than on programming.  A clear description of the physics algorithm naturally allows to communicate analyses easily between different groups, such as between experimentalists and phenomenologists, or professional physicists with students or interested public.  ADL and tools, in particular CutLang, are already being employed in several physics studies as well as for training students at different levels.  They are also being adapted for use with LHC Open Data.  In the following, we will introduce basics of ADL and {\tt CutLang}, and present current use cases for education and outreach.

\section{ADL and CutLang}

The main focus of ADL is event processing: it describes definition of objects (jets, electrons, muons, etc.), variables (e.g. invariant mass) and event selections (e.g. signal, control or validation regions).  ADL consists of a plain text file called the ADL file, where the analysis physics algorithm is described with the ADL syntax.  The ADL file is accompanied by a library of self-contained functions encapsulating variables non-trivial to express with the ADL syntax such as complex kinematic variables (e.g. aplanarity, stransverse mass $M_{T2}$) or non-analytic variables (e.g. machine learning discriminants, numerical descriptions of efficiencies).  The ADL file consists of blocks separating object, variable and event selection definitions. Blocks have a keyword-expression structure, where keywords specify analysis concepts and operations.  Syntax includes mathematical and logical operations, comparison and optimization operators, reducers, 4-vector algebra and standard functions used in an analysis (e.g. $\delta\phi$, $\delta R$).  Histograms can also be defined using ADL.  Moreover, pre-existing selection results including counts and uncertainties (e.g. from publications) can be documented in the ADL file.  Work is ongoing to incorporate systematic uncertainties.  More detailed information on the ADL syntax can be found in~\cite{Unel:2021edl}.

\begin{table}[htbp]
\caption{Description of a simple analysis (inspired by supersymmetry searches) using the ADL syntax.  This description can be directly processed with CutLang over events.}
\label{tab:ADLdesc}
\begin{minipage}{0.33\linewidth}
\fontsize{7 pt}{0.85 em}
\selectfont
\begin{verbatim}
# OBJECTS
object goodJet
  take jet
  select pT(jet) > 30
  select abs(eta(jet)) < 2.4

object goodMuon
  take Muon
  select pT(Muon) > 30
  select abs(eta(Muon)) < 2.4

object goodEle
  take Ele
  select pT(Ele) > 30
  select abs(eta(Ele)) < 2.5

object goodLep 
  take union(goodEle, goodMuo)
\end{verbatim}
\end{minipage}
\begin{minipage}{0.66\linewidth}
\fontsize{7 pt}{0.85 em}
\selectfont
\begin{verbatim}
# EVENT VARIABLES
define HT = fHT(jets)
define MTl = Sqrt( 2*pT(goodLep[0]) * MET*(1-cos(phi(METLV[0]) - phi(goodLep[0]) )))

# EVENT SELECTION
region SR
  select size(jets) >= 2
  select HT > 200
  select MET > 200
  select MET / HT <= 1
  select Size(goodEle) == 0
  select Size(goodMuon) == 0
  select dphi(METLV[0], jets[0]) > 0.5
  select dphi(METLV[0], jets[1]) > 0.5
  select size(jets) >= 3 ? dphi(METLV[0], jets[2]) > 0.3 : ALL
  select size(jets) >= 4 ? dphi(METLV[0], jets[3]) > 0.3 : ALL
  histo hMET , "met (GeV)", 40, 200, 1200, MET
  histo hHT , "HT (GeV)", 40, 200, 1600, HT
\end{verbatim}
\end{minipage}
\end{table}

\begin{figure}
    \centering
    \includegraphics[width=0.6\textwidth]{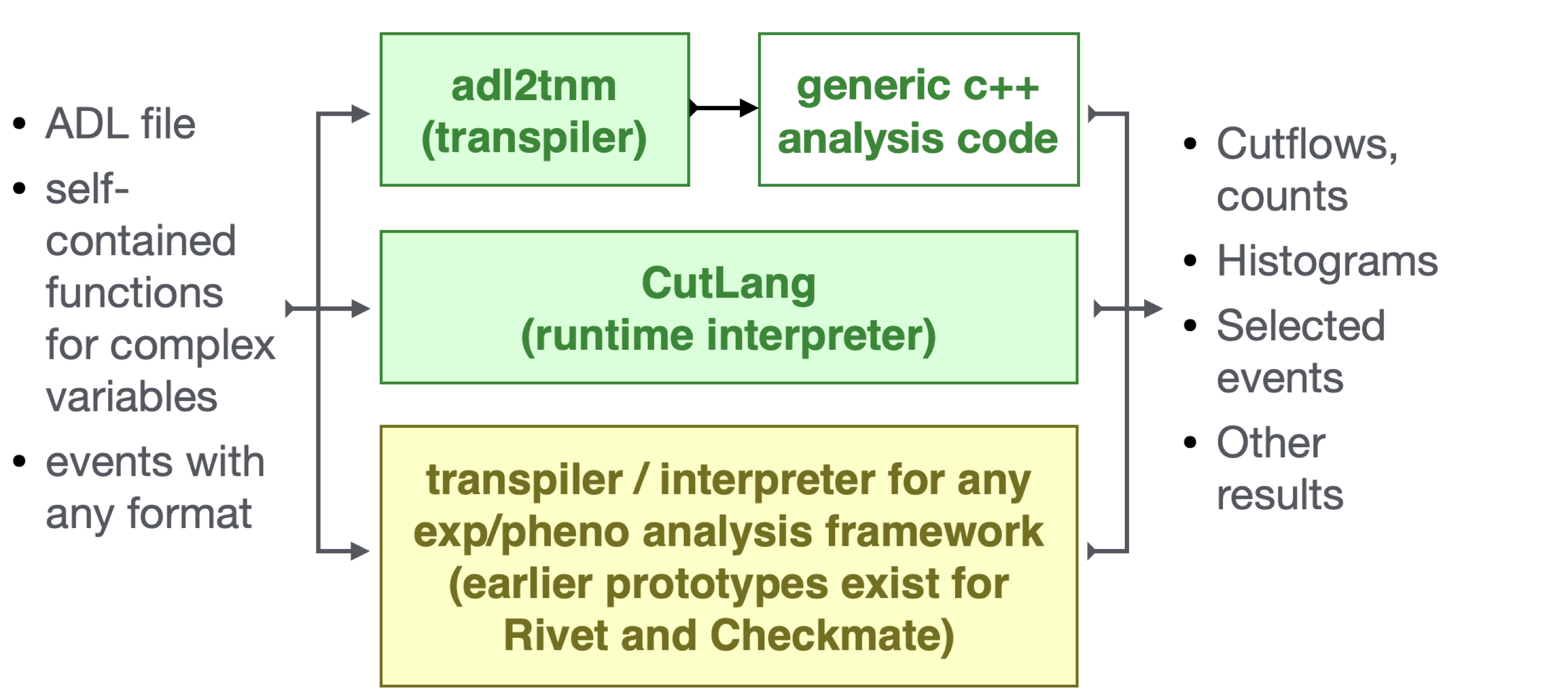}
    \caption{ADL analyses flow with different tools.  Inputs to and outputs from a typical ADL analysis.}
    \label{fig:ADLtools}
\end{figure}

An analysis written in ADL can be run with any framework that parses and executes the ADL syntax.  Figure~\ref{fig:ADLtools} shows a typical ADL analysis workflow.  Input to the analysis are the ADL file, external functions (if relevant) and events stored in CERN ROOT analysis framework ntuples.  Analysis outputs are cutflows, counts, histograms, and if needed, sets of selected events.  

Currently CutLang~\cite{Sekmen:2018ehb, Unel:2019reo, Unel:2021edl} is the most developed tool for ADL analyses. CutLang is a {\emph runtime interpreter} meaning that it can directly run on the ADL file without requiring compilation (except for once during installation) or extra coding by the user.  CutLang itself is based on {\tt C++} and ROOT.  It performs automatic ADL parsing by the state-of-the-art tools Lex \& Yacc.  The CutLang framework can work with multiple input event formats regularly used in HEP, such as Delphes fast simulation, CMS NanoAOD, ATLAS/CMS Open Data, LVL0, FCC, while more formats can be easily added.  For analyses with simple object definitions not requiring complex identification criteria, CutLang can run the same ADL file on different input types.  This is particularly useful for education purposes, since training exercises can combine signal and background processes from different sources (e.g. backgrounds from CMS Open Data and signals from Delphes).   CutLang produces output in ROOT files, which contains cutflows, bins and histograms for each event selection region in a separate TDirectory, as well as the ADL description itself for provenance tracking.

\begin{wraptable}{r}{8cm}
\caption{Computing environments in which CutLang is available for various operating systems.}
\begin{center}
\fontsize{9 pt}{0.85 em}
\selectfont
\begin{tabular}{|c|c|c|c|}
\hline
Environment & Linux & MacOS & Windows \\
\hline
Native & \checkmark & \checkmark &  \\
Docker  & \checkmark & \checkmark & \checkmark \\
Conda commandline  & \checkmark & \checkmark &  \\
Conda / Jupyter  & \checkmark & \checkmark &  \\
Jupyter web  & \checkmark & \checkmark & \checkmark \\
\hline
\end{tabular}
\end{center}
\label{tab:CLenv}
\end{wraptable}%

\begin{figure}
    \centering
    \includegraphics[width=0.8\textwidth]{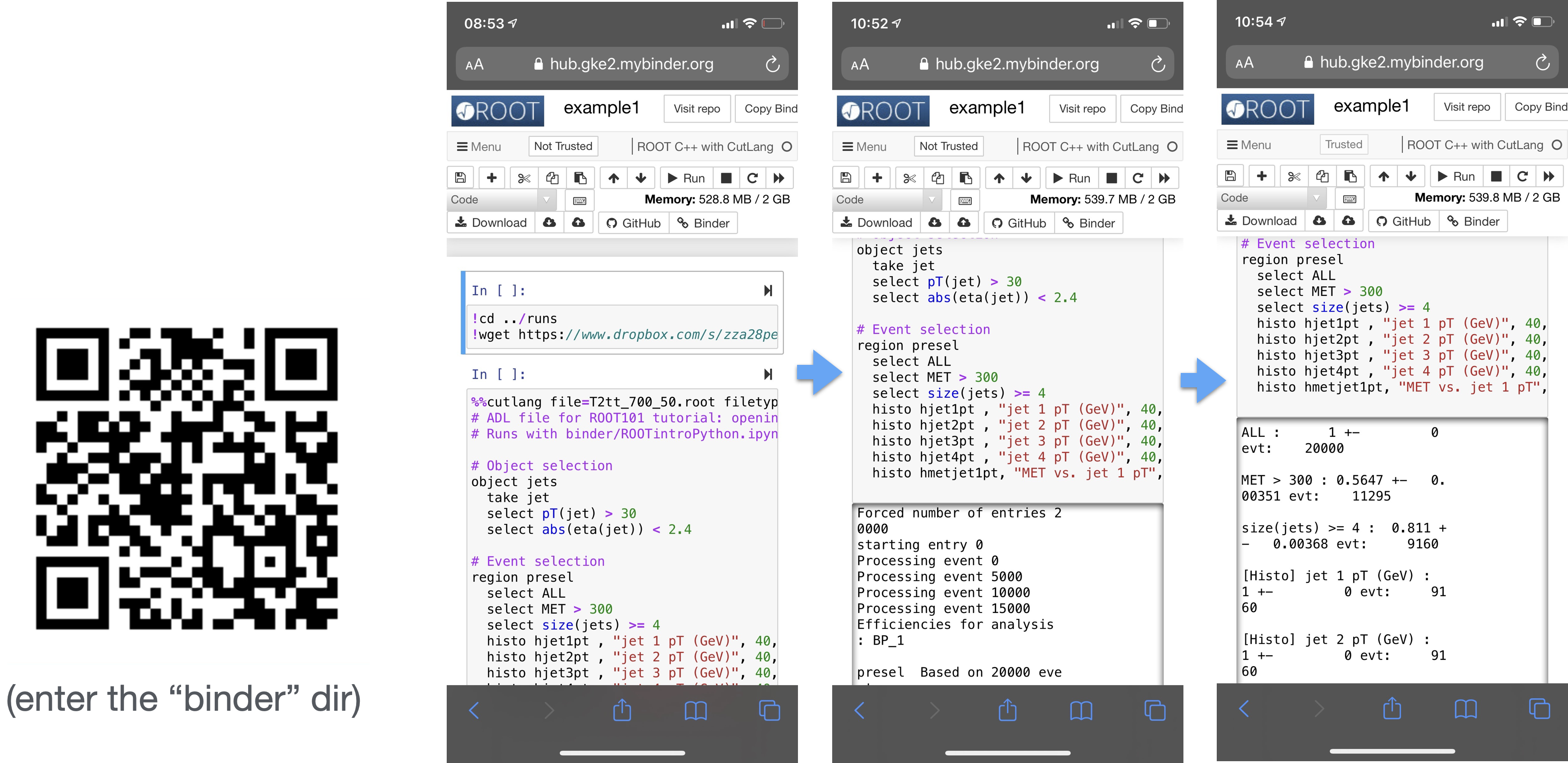}
    \caption{Running an ADL analysis with CutLang on a mobile phone via Jupyter / binder interface.  The interface can be accessed by the QR code.}
    \label{fig:CLbindermobile}
\end{figure}

To reach the widest possible audience, CutLang is adapted to run on multiple computing environments as shown in Table~\ref{tab:CLenv}. Docker and Conda environments enable directly working with a preinstalled version of CutLang, and thus to bypass all potential issues that could be encountered during initial setup and compilation in native environments.  Recently, a CutLang Jupyter kernel was developed, that allows to execute ADL directly in Jupyter notebooks.  User simply enters the analysis written in ADL syntax directly in a Jupyter notebook cell along with input and output file names, input data format and number of events to run, and executes the cell.  Jupyter can be run via Conda, but also directly on a web browser via binder, where the latter only requires access to a web browser, which makes this setup an excellent tool for short outreach exercises.  The jupyter / binder interface can even be accessed from a mobile phone through a link redirected by a QR code, as shown on Figure~\ref{fig:CLbindermobile}.  This feature can be diversely exploited for outreach activities.

\section{Education and outreach with ADL/CutLang}

ADL and CutLang are used for implementing various physics analyses for experimental and phenomenological studies~\cite{Paul:2020mul, Brooijmans:2020yij}.  A growing LHC ADL analysis database is available in GitHub~\cite{adllhcanalyses}, containing both complete and simplified analyses that can be  run by CutLang.  This database is a first step towards long term analysis preservation.  It also serves as a learning database presenting analysis concepts to physics students, high school students, or anyone willing to learn.

ADL and CutLang were recently used in two schools for physics students.  First one was the 5th School of Computing Applications in Particle Physics, organized on 3-7 Feb 2020 in Istanbul, Turkey, dedicated to training an audience of mostly undergraduate students on LHC analysis with ADL/CutLang.  Students had no or little experience in programming or analysis.  After 3 days of lectures on LHC physics and short exercises with CutLang, students performed 1.5 day long exercises in groups of 3-4 people.  Each group was assigned an ATLAS or CMS analysis, with the task to implement the analysis in ADL, run it with CutLang over MC and get simple results. The MC signal and background events were either generated by the facilitators or obtained from ATLAS/CMS Open Data portal.  The students were able to perform object and signal region selections and draw plots.  The school's program and analysis exercises are detailed in~\cite{Adiguzel:2020brl}.  The second was the Vietnam School of Physics (VSOP-26), organized on Nov 29-Dec 11 2020 in Guy Nhon, Vietnam.  This was a general particle physics school for undergraduate or junior graduate students with theory background and no coding or analysis experience.  Three hours of hands-on exercises were given using the Jupyter / binder interface, whose instructions can be found in~\cite{}.  Students were able to write simple analyses for basic Z boson reconstruction and an event selection for a simple supersymmetry search.    

ATLAS and CMS are publishing increasing amounts of Open Data (OD) consisting of real or simulated events for use in academic research or education. Example analysis codes are provided by the OD teams but entry level to analysis is difficult for non-professionals.  ADL and CutLang can facilitate OD analysis at all levels, and are already adapted for this purpose.  ATLAS has published Run1 and Run2 OD, mostly for education purposes, and CutLang can process these OD formats.  ADL descriptions for simple $Z \rightarrow \ell \ell$ and $H \rightarrow \gamma \gamma$ analyses are available as examples.  On the other side, CMS published OD  for research in AOD format, which requires CMS software to run, and for education in simple ntuple format, which can already be processed by CutLang.  In the near future, CMS will release data in the NanoAOD format, a simple but inclusive ntuple format that can be used for research and education.  CutLang can already process NanoAOD, though refinements are ongoing to provide analysis capabilities at every level of detail.  Simple analysis examples exist for Z and Higgs boson reconstruction, and a demo was presented at the CMS OD4Theorists Workshop on 2020 based on the more detailed CMS $H \rightarrow \tau \tau$ analysis.  Discussions are ongoing with both ATLAS and CMS OD teams to establish ADL/CutLang as an accessible OD analysis model.

\section{Summary and outlook}

ADL/CutLang is a promising and feasible approach for practical LHC data analysis, applicable both to research and education.  It decouples physics logic and algorithm from frameworks, thereby allowing the focus to be on analysis design rather than on programming technicalities.  ADL/CutLang does not require any system or software level expertise to run analyses on data, making it easier to explore ideas and ask ``what if" questions of data.  ADL and CutLang are under constant development, and several experimental and phenomenological studies are in progress using this approach.  The ADL analysis database is expanded with more physics analysis examples, which can serve both academic and education purposes.  Data analysis schools with ADL/CutLang planned in Korea, Turkey and USA.  A complete integration with LHC Open Data is in progress in collaboration with the Open Data teams.  Additionally, an education module for high school students and teachers based on ADL/CutLang and Open Data is being designed.  All information and developments are systematically documented in the ADL web portal \href{cern.ch/adl}{cern.ch/adl}.  With the low barrier to entry ADL/CutLang promises, we hope this approach will make LHC data analysis more accessible and broaden the pool of LHC data explorers.

\vspace{0.5cm}

\noindent {\bf Acknowledgements:} We thank our collaborators in the growing ADL/CutLang team for their great efforts.  We also thank the ATLAS and CMS Open Data teams for useful discussions and feedback.  Work of SS on ADL/CutLang is supported by the National Research Foundation of Korea (NRF), funded by the Ministry of Science \& ICT under contract 2021R1I1A3048138.



\begin{thebibliography}{99}


\bibitem{Brooijmans:2016vro}
G.~Brooijmans, {\it et. al.}
[arXiv:1605.02684 [hep-ph]].

\bibitem{Brooijmans:2018xbu}
G.~Brooijmans, {\it et. al.}
[arXiv:1803.10379 [hep-ph]].

\bibitem{Sekmen:2018ehb}
S.~Sekmen and G.~Unel,
Comput. Phys. Commun. \textbf{233} (2018), 215-236
doi:10.1016/j.cpc.2018.06.023
[arXiv:1801.05727 [hep-ph]].

\bibitem{Unel:2019reo}
G.~Unel, S.~Sekmen and A.~M.~Toon,
J. Phys. Conf. Ser. \textbf{1525} (2020) no.1, 012025
doi:10.1088/1742-6596/1525/1/012025
[arXiv:1909.10621 [hep-ph]].

\bibitem{Unel:2021edl}
G.~Unel, S.~Sekmen, A.~M.~Toon, B.~Gokturk, B.~Orgen, A.~Paul, N.~Ravel and J.~Setpal,
Front. Big Data 4:659986
doi:10.3389/fdata.2021.659986
[arXiv:2101.09031 [hep-ph]].


\bibitem{Paul:2020mul}
A.~Paul, S.~Sekmen and G.~Unel,
Eur. Phys. J. C \textbf{81} (2021) no.3, 214
doi:10.1140/epjc/s10052-021-08982-4
[arXiv:2006.10149 [hep-ph]].

\bibitem{Brooijmans:2020yij}
G.~Brooijmans, {\it et. al.}
[arXiv:2002.12220 [hep-ph]].

\bibitem{adllhcanalyses}
ADL Analysis Database: https://github.com/ADL4HEP/ADLLHCanalyses


\bibitem{Adiguzel:2020brl} A.~Ad\i{}g\"uzel, O.~\c{C}ak\i{}r, \"U.~Kaya, V.~E.~\"Ozcan, S.~\"Ozt\"urk, S.~Sekmen, I.~Turk Cakir and G.~\"Unel, 
Eur. J. Phys. \textbf{42} (2021), 035802 
doi:10.1088/1361-6404/abdf67 
[arXiv:2008.12034 [hep-ph]]. 



\end{thebibliography}
\end{document}